\documentclass[12pt]{article}
\usepackage{amsmath,amssymb,epsf
}
\newcommand{\beq}{\begin{equation}}
\newcommand{\eeq}{\end{equation}}
\newcommand{\ba}{\begin{array}}
\newcommand{\ea}{\end{array}}
\newcommand{\bea}{\begin{eqnarray}}
\newcommand{\eea}{\end{eqnarray}}
\newcommand{\bean}{\begin{eqnarray*}}
\newcommand{\eean}{\end{eqnarray*}}

\newtheorem{theorem}{Theorem}[section]
\newtheorem{prop}[theorem]{Proposition}

\newtheorem{exe}[theorem]{Exercise}

\newtheorem{remark}[theorem]{Remark}

\newtheorem{proof}{Proof.}
\newenvironment{pf}{\begin{proof} \rm}{\hfill$\square$ \end{proof}}
\newcommand{\CH}{{\cal H}}

\newcommand{\CE}{{\cal E}}
\newcommand{\CW}{{\cal W}}

\newcommand{\CF}{{\cal F}}
\newcommand{\CM}{{\cal M}}

\newcommand{\CU}{{\cal U}}
\newcommand{\CV}{{\cal V}}
\makeatletter
\@addtoreset{equation}{section}

\makeatother

\newcommand{\CC}{{\mathbb C}}

\def\la{\lambda}

\newcommand{\cmp}[3]{Comm. Math. Phys. {\bf #1} (#2), #3}

\newcommand{\prl}[3]{Phys. Rev. Lett. {\bf #1} (#2), #3}

\newcommand{\lmp}[3]{Lett. Math. Phys. {\bf #1} (#2), #3}

\newcommand{\rref}[1]{(\ref{#1})} 

\def\dsl{\displaystyle}

\newcommand{\del}{{\partial}}

\def\mat2#1#2#3#4{{\left(\begin{array}{cc}#1 & #2\\ #3 & #4 \end{array}\right)}}
\def\mats2#1#2#3#4{{\left(\begin{array}{cc}#1 & #2\vspace{2truemm} \\ #3 & #4
\end{array}\right)}}
\def\vec2#1#2{{\left(\begin{array}{c}
#1 \\ #2 \end{array}\right)}}

\def\ddd#1#2{\displaystyle{\frac{\partial #1}{\partial #2}}}

\def\fddd#1#2{\displaystyle{\frac{\delta #1}{\delta #2}}}
\newcommand{\dddx}{{\ddd{}{x}}}

\def\alg{{\mathfrak g}}

\def\ger{hierarch}

\def\bih{bihamiltonian}

\def\ger{hierarch}

\newcommand{\expo}[1]{{e^{#1}}}
\newcommand{\wid}[1]{\widehat{#1}}

\newcommand{\ddt}{{\frac{d}{dt}}}
\begin{document}
\null
\begin{flushright}
Ref. SISSA 35/2005/FM
\end{flushright}
\vspace{0.6truecm} \baselineskip=24pt
\begin{center}
{\Large\bf 
On a Camassa-Holm type equation\\
with two dependent variables}
\end{center}
\vspace{0.3truecm}
\begin{center}
{\large
Gregorio Falqui\\ \vspace{0.2 truecm}
SISSA, Via Beirut 2/4, I-34014 Trieste, Italy\\
}
\end{center}
\baselineskip=18pt \vspace{0.5truecm} \noindent{ {\bf Abstract}:
We consider a generalization of the Camassa Holm (CH) equation
with two dependent variables, called CH2, introduced in
\cite{LZ04}. We briefly provide an alternative derivation of it
based on the theory of Hamiltonian structures on (the dual of) a
Lie Algebra. The Lie Algebra here involved is the same algebra
underlying the NLS \ger y. We study the structural properties of
the CH2 \ger y within the \bih\ theory of integrable PDEs, and
provide its Lax representation. Then we explicitly discuss how to
construct classes of solutions, both of peakon and of
algebro-geometrical type. We finally sketch the construction of a
class of singular solutions, defined by setting to zero
one of the two dependent variables.}
\section{Introduction}
The relevance of Camassa Holm equation, first discovered by means
of geometric considerations by Fokas and Fuchssteiner \cite{FF81},
was brought to the light in \cite{CH93}, where it was obtained as
a suitable limit of the Green-Naghvi equations. One of its more
interesting features is that it is an integrable approximation of
an order higher than KdV to the Euler equations in one spatial
dimension.

From the mathematical point of view, until quite recently the CH
\ger y was possibly the only well known example of integrable \ger
y not comprised in the Dubrovin-Zhang classification
scheme of evolutionary \bih\ hierarchies \cite{DZ02}.
The reason for this is
that the CH \ger y does not admit a formulation by means of a
$\tau$-function. This lacking is reflected in the properties of
notable classes of solutions. Indeed, bounded traveling waves for
the CH equation (termed {\em peakons}) develop a discontinuity in
the first derivatives, and the evolution properties of finite
gap solutions, as discussed in \cite{achm94,
  CMc99,acfhm01}, are somewhat peculiar, even if
they can be expressed in terms of hyperelliptic curves
as in the KdV case. The Whitham modulation theory
associated with genus one solutions of CH, discussed in
\cite{AG05}, also presents some non standard features.

The integrable equations we are going to discuss in the present
paper are conservation laws for two dependent variables of the
form:
\begin{equation}
  \label{eq:ch2-eq}
\begin{array}{lcl}
{(v- v_{x})_{t}}
&=& (-2\rho
    u+v^2- vv_x)_x\\
    {(u+ u_{x})_{t}}&=&(2uv+ u_x v)_x,
\end{array}
\end{equation}
with $\rho$ a parameter. These equations was derived (for
$\rho=1$) in \cite{LZ04} within the framework of the general
deformation theory of hydrodynamic \ger ies of \bih\ evolutionary
PDEs \cite{DZ02}; more recently, a related equation has been
considered in \cite{CLZ05}, within the framework of reciprocal
transformations, and the properties of solitary waves and
2-particle like solutions described.

Actually, a similar equation was introduced by Olver and Rosenau
in \cite{OR95}, as a deformation of the Boussinesq system;
traveling wave solutions of the resulting equation can be found in
\cite{LOR99}.
 In particular, in the paper \cite{OR95}, various non-standard
integrable equations were defined. The key observation was that
one can define \bih\ pencils from a given Poisson tensor using
scaling arguments, and hence apply a recipe used in \cite{FF81}.

Our derivation of the equation resembles the one of \cite{OR95};
however, on the one hand we will take advantage of the fact that the
phase space is the dual of a Lie algebra, and, on the other
hand, we will require that the hydrodynamical limit of the resulting equations
be ``substantially'' the same as that of the classical equation we
are starting from. As it will be briefly sketched in Section \ref{sec:2}, the
latter are the well known AKNS equations, that is, a complex form
of the Nonlinear Schr\"odinger equations.

In the core of the paper we will study both formal and concrete
properties of the \ger y \rref{eq:ch2-eq},
which we call {\em CH2 \ger y}. At first
we  will characterize it from the \bih\ point of view, find its Lax
representation, and discuss the
associated {\em negative} \ger y.

Then we will discuss the features of two main classes of solutions. We
will show that it admits peaked traveling waves, and show that
these solutions can be consistently superposed, giving rise to the same
finite dimensional Hamiltonian system associated with the N-peakon solution of
the CH equation. Then we will address the problem of characterizing
algebro-geometric solutions. We will show that the same phenomenon that happens
in the Harry-Dym and CH finite gap solutions,  namely the occurrence of
non-standard Abel-Jacobi maps, is reproduced here.

We will end noticing that the reduction of the CH2 equations on the
submanifold $u(x,t)=0$, where they yield
the one-field equation
\[
(v-v_x)_t=(v^2-vv_x)_x
\]
admits weak solutions of traveling wave type
with peculiar interaction properties.

\section{Some remarks on the CH \ger y}\label{sec:1}
We  herewith collect some remarks on the CH equation
\begin{equation}
  \label{eq:che}
  u_t-u_{xxt}=-3u u_x+2 u_x u_{xx}+u u_{xxx},
\end{equation}
and highlight those features that will lead us to define its
$2$-field generalization the next Sections. In particular, we will
focus on a connection of CH with the KdV equation $
u_t=-u_{xxx}+6uu_x$ which was (although implicitly) pointed out
\cite{KM00} in the framework  of Euler equations on
diffeomorphisms groups.

From the \bih\ point of view, the KdV theory can be regarded as the
Gel'fand--Zakharevich \cite{GZ93} theory of the Poisson pencil
\begin{equation}
  \label{eq:ppkdv}
Q-\la P=-\del_x^3+2u\del_x+u_x-\la \del_x.
\end{equation}
defined on a suitable space $\CU$ of functions $u=u(x)$ of one
independent variable. As it has been long known, $\CU$ admits a
very natural geometrical interpretation of dual of the Witt
algebra $\CV$ of vector fields in ``one dimension''\footnote{E. g., on the
punctured plane $\CC^*$.}, whose elements will be represented with
$f\dsl{\ddd{}{x}}$, endowed with the natural Lie bracket. Actually,
a closer look at the expression of the KdV Poisson pencil
\ref{eq:ppkdv} shows that, calling
\begin{equation}
  \label{eq:ppp}
  Q_\omega=\del^3_x,\quad P=\del_x,\quad Q_h=2u\del_x+u_x,
\end{equation}
we are facing a {\em triple} of Poisson tensors, $P,Q_\omega,
Q_h$, all of which have a well defined geometrical and algebraic
meaning. Namely:
\begin{enumerate}
\item $Q_h$ is the Lie Poisson structure on $\CV^*$. \item $P$ is
associated with the coboundary ${c}(f\dddx,g\dddx)=\oint fg_x$.
\item $Q_\omega$  is associated with the Gel'fand-Fuchs cocycle
$\omega(f\dddx,g\dddx))=\oint f g_{xxx}$.
\end{enumerate}
So
 $(\CV^*, \{Q_h,Q_\omega,P\})$ is actually a `tri'--Hamiltonian manifold,
that is, that for any (complex) numbers $\eta, \la,\mu$ the linear
combination
\begin{equation}
  \label{eq:poitri}
P_{\eta, \la,\mu}=\eta Q_h+\mu Q_\omega+\la P
\end{equation}
is a Poisson tensor. This  follows from the obvious compatibility
between the two constant tensors $Q_\omega$ and $P$, and the known
property that the compatibility condition between the Lie Poisson
structure and any constant structure on the dual of a Lie algebra
$\alg$ coincides with the closure condition for 2-cochains in the
cohomology  of $\alg$.

In this respect we see that KdV can be regarded as the GZ theory
on $\CV^*$, equipped with a particular pencil extracted from the
web of Poisson tensors~\rref{eq:poitri}, namely that obtained
``freezing'' (in the parlance of \cite{KM00}) $\mu$ and $\eta$ to
the value $1$.

The route towards the (bi)Hamiltonian approach to the CH equation
is similar. Indeed, we can (following \cite{CH93}) consider, in
the space $\CM$ defined by the associated dependent variable (the
momentum) $m=u-u_{xx}$, the two Poisson structures
\begin{equation}
  \label{eq:ppch}
  Q_h=2m\del_x+m_x,\quad  \text{  and } R=Q_\omega-P=\del^3-\del=-(\del_x
\cdot({\mathbf 1}-\del_x^2)).
\end{equation}
They form another Poisson pencil extracted from the web
\rref{eq:poitri}, and provide the CH equation with a
\bih\ representation,
with Hamiltonian functions given by
\begin{equation}
  \label{eq:ch-ham2}
  H_1=\frac12\int u^2 dx, \quad H_2=\frac12\int (u^3+uu_x^2)\, dx.
\end{equation}
We finally recall the following points:
\begin{description}
\item[i)] Apart from an inessential multiplicative factor, the
hydrodynamical limit of CH and KdV coincide with the Burgers
equation $u_t=u u_x.$ \item[ii)] One can define a {\em negative CH}
\ger y by iteration of the Casimir $H_{-1}=\int\sqrt{m}dx$ of $Q_h$.
 \item[iii)] The traveling wave solution of the CH equation is
the {\em peakon}
\begin{equation}
  \label{eq:peak}
  u(x,t)=c\exp{(-|x-ct|)};
\end{equation}
More generally, the CH equations admit $N$-peakon solutions of the
form
\begin{equation}
  \label{eq:npeak}
  u(x,t)=\sum p_i(t)\exp{(-|x-q_i(t)|)},
\end{equation}
where $\big(p_i(t),q_i(t)\big)$ describe the geodesic motion on a manifold
with (inverse) metric $ g^{i,j}=\exp(-|q_i-q_j|)$ \item[iv)] The
CH equation admits a Lax pair, being the compatibility condition
of the linear system
\begin{equation}
  \label{eq:laxpair}
  \psi_{xx}=\big(\frac14-\frac{m}{2\la}\big)\psi,\quad
\psi_t=-(\la+u)\psi_x+\frac12 u_x\psi.
\end{equation}
\item[v)] Cnoidal waves for CH satisfy the following relation
tying $u'=\frac{d}{dr}u(r)$ and $u=u(r)$, where $r=x-ct$:
\begin{equation}
  \label{eq:cnwav}
  {u'}^2(c-u)+u^3+a_2 u^2+a_1 u +a_0=0,
\end{equation}
that is, they involve the integration of an Abelian differential
of the third kind on this genus $1$ curve, having simple poles at
$u=\pm\infty$. The same holds \cite{achm94,acfhm01} for algebro
geometrical solutions of higher genus.
\end{description}

\section{From the AKNS hierarchy to the CH2 hierarchy}\label{sec:2}
The AKNS equations describe isospectral deformations of the linear
operator
\[
L=\del_x-\left( \begin{array}{cc} \la& p(x)\\
    q(x)&-\la\end{array}\right),
\]
and read:
\begin{equation}\label{eq:akns}
\dot p=-p_{xx}+qp^2\qquad \dot q=q_{xx}+q^2p.
\end{equation}
As it is well known, these equation admit a \bih\ formulation, and
their restriction to $p(x)=\bar{q}(x)$ yields (after setting $t\to
it$) the Non-Linear Schr\"odinger equation
\begin{equation}
  \label{eq:nls}
  -i\dot q=q_{xx}+|q|^2q.
\end{equation}
 A perhaps less known fact is the following. Under the
``coordinate change''
\begin{equation}
  \label{eq:coch}
  v=q_x/q,\quad u=p\,q,
\end{equation}
the AKNS equations become
\begin{equation}
  \label{eq:19}
  u_t={-u_{xx}+2(uv)_x},\quad
  v_t={v_{xx}+2vv_x-2u_x}.
\end{equation}
They admit a \bih\ formulation via the Poisson pair
\begin{equation}
  \label{eq:newpb}
  P_0=\mat2{0}{\del_x}{\del_x}{0},\quad P_1=\mat2{2u\del_x +u_x}{-\del_x^2+v\del_x}{\del_x^2+\del_x\,v}{-2\del_x},
\end{equation}
and Hamilton functions
\begin{equation}
h_1=\int_{S^1} (uv) dx, \quad h_2=\int_{S^1} (uv^2-u^2-u_x v) dx.
\end{equation}

In analogy with the KdV case, the brackets \rref{eq:newpb}, being
at most linear in $(u,v)$, admit a sound Lie-algebraic
interpretation.

Let $\CW=\CV\ltimes\CF$ be the semidirect product of the algebra
of vector fields on $\CC^*$ with the algebra $\CF$ of functions on
$\CC^*$, that is the space of pairs
\begin{equation}
(f\dddx, g),\quad f, g\in \CC((x)),
\end{equation}
equipped with the Lie bracket
\[
[(f_1\dddx, g_1),(f_2\dddx, g_2)]=\big((f_1f_{2,x}-f_2f_{1,x})
\dddx, f_1g_{2,x}-f_2 g_{1,x}\big).
\]
Decomposing the second Poisson tensor of Eq. \rref{eq:newpb} as
$P_1=P_1^{h}+P_1^{c}+P_1^{lp}$, where:
\begin{equation}
  \label{eq:p1dec}
  P_1^{h}=\mat2{0}{0}{0}{-2\del_x},\>
  P_1^{c}=\mat2{0}{-\del^2_x)}{\del^2_x}{0},\>
P_1^{lp}\mat2{2u\del_x +u_x}{v\del_x}
{\del_x v}{0},
\end{equation}
we see:
\begin{description}
\item[i)] $P_0$ is the bracket associated with the coboundary $
\int(f_1g_{2,x}-f_2g_{1,x})$. \item[ii)] $P_1^{lp}$ is the Lie
Poisson bracket on $\CW^*$. \item[iii)] $P_1^h$ is (a multiple of)
the bracket associated with the Heisenberg cocycle $\omega^h:=\int
g_{1,x} g_2$. \item[iv)] $P_1^c$ is the bracket associated with
the non-trivial cocycle $ \int (f_1g_{2,xx}-f_2g_{1,xx})$.
\end{description}
These statements can be proven referring to, e.g., \cite{ADKP86},
where  $H^2(\CW)$ was computed to be three--dimensional, with
generators given by $\omega^c,\omega^h$ and the  Virasoro cocycle
$ \omega^v=\int f_1f_{2,xxx}$. This tensor does not enter the
local representation of the AKNS/NLS theory we are dealing with,
but rather the dispersive generalization of the Boussinesq system
\begin{equation}\label{boh2}
u_t=(u^2+h+u_x)_x\quad
h_t=(2uh+u_{xx}-h_x)_x
\end{equation}
discussed in \cite{Kup95}.

From the theory of Poisson brackets on (duals of) Lie algebras we
get:
\begin{prop}
The linear combination
\[
P(\la_1,\la_2,\la_3,\la_4,\la_5)=\la_1P_0+\la_2P_1^{lp}+\la_3P_1^h+\la_4
P_1^c+\la_5 P^{vir}
\]
is, for all values of the parameters $\la_i$, a Poisson tensor.
\end{prop}
In the sequel we will study the pencil $P_\la$ defined by:
\begin{equation}
  \label{eq:pencil}
  P_\la=(P^0-P^c)-\la (\rho P_1^h+P_1^{lp})\equiv P-\la Q,
\end{equation}
where $\rho$ is a (fixed) parameter.
Clearly enough, different
choices could be made. For example, in \cite{OR95}, the pair
$P'=P^0+P^c+P^h$, $Q'=P^{lp}$ has been considered, on the basis of scaling
considerations.
With our choice, the dispersionless limits of \rref{eq:pencil} and of
\rref{eq:newpb} coincide (up to the parameter $\rho$). Essentially we once
again mimicked the Camassa-Holm case in moving the ``dispersive cocycle''
$P^c$ to the first member of the Poisson pair.

We notice that the Poisson tensor $P$ of Eq. \rref{eq:pencil}
admits, in analogy with the CH case, the factorization
\[
P=\del_x\cdot \mat2{0}{\mathbf{1}-\del_x}{\mathbf{1}+\del_x}{0}\equiv
\del_x\cdot \Psi
\]
in which the operator
$\Psi$
is {\em symmetric}. This will enable us to write evolutionary PDEs for the
``physical'' dependent variables $u,v$ by studying the
Lenard-Magri sequences associated with the Poisson pencil $Q-\la
P$ defined on the space of the associated dependent variables
\[
\eta=v- v_x,\quad \xi=u+u_x,
\] as in the ordinary CH case, at least for
the first few steps.
In matrix form, the
Hamiltonian vector fields associated with the Poisson pencil
$P_\la$ of \rref{eq:pencil} will have the following explicit expression:
\begin{equation}
  \label{eq:mapenc}
  \vec2{{\eta}_t}{{\xi}_t}=\left(\mat2{-2\rho\del_x}{\del_x\cdot\eta} {\eta\cdot\del_x}{2\xi\del_x
+\xi_x} -\la \mat2{0}{\del_x-\del^2_x}{\del_x+\del^2_x}{0}\right)
\cdot \vec2{\fddd{H}{\eta}}{\fddd{H}{\xi}}.
\end{equation}
\subsection{The \ger y}\label{subs:3.1} To write the
analogue of the CH equations we apply the Gel'fand-Zakharevich
iteration scheme to the Poisson pencil $Q-\la P$ defined by
\rref{eq:mapenc}. In particular, starting with the Casimir\footnote{The other
  regular Casimir $\int \eta$ of $P$ is a common Casimir.} $
H_1=\int \xi\, dx \equiv \int u\, dx$ the first vector field of
the (``positive'') \ger y is $x$-translation.

This vector field is Hamiltonian with respect to $P$ as well, with
Hamiltonian function $H_2$ given by
\begin{equation}
  \label{eq:H1}
  H_2=\int u\eta\, dx=\int u(v-v_x)\, dx.
\end{equation}
The equation we are interested in, (the {\em CH2 equation}),
is the next equation in the
Lenard-Magri ladder, that is
\begin{equation}
  \label{eq:zlfch-eq}
  \vec2{\ddd{\eta}{t}}{\ddd{\xi}{t}}=Q\cdot
\vec2{\fddd{H_2}{\eta}}{\fddd{H_2}{\xi}}\Leftrightarrow
\begin{array}{lcl}
{(v- v_{x})_{t}}
&=& \del_x(-2\rho
    u+v^2- vv_x)\\
    {(u+ u_{x})_{t}}&=&\del_x(2uv+ u_x v).
\end{array}
\end{equation}
 This equation is actually \bih. Indeed, it is easily verified that it is
 Hamiltonian w.r.t. $P$, with (second) Hamiltonian given by:
 \begin{equation}
   \label{eq:scndH}
   H_3=\int \big(-\rho u^2+u(v^2- vv_x)\big)\, dx.
 \end{equation}
The Lenard-Magri recursion could be prolonged further. As in the CH
case, the next Hamiltonians are no more local functional in $(u,v)$.

Along with the positive \ger y we can define a {\em negative} \ger
y. We switch the roles of $P$ and $Q$, that is, we seek for a
Casimir function for the pencil $P-\la Q$. Actually, as a first
step we seek for the differential of this Casimir, that is for two
Laurent series
$\beta_\eta=\beta_\eta^0+\frac1\la\beta_\eta^1+\cdots,
\beta_\xi=\beta_\xi^0+\frac1\la\beta_\xi^1+\cdots$ satisfying
\begin{equation}
  \label{eq:caspoi}
\mat2{2\rho\la\del_x}{\del_x-\del_x^2-\la\del_x\cdot\eta}
{\del_x+\del_x^2-\la \eta\del_x}{-\la(2\xi\del_x
  +\xi_x)}\cdot\vec2{\beta_\eta}{\beta_\xi}=\vec2{0}{0}.
\end{equation}

We see that these two one-forms must satisfy the equations:
\begin{equation}
  \label{eq:casint}
  \begin{split}
&2\rho\la\beta_\eta+\beta_\xi(1-\la\eta)-\beta_{\xi,x}=F_1(\la)\\
&\la(\rho\beta_\eta^2-\beta_\eta\beta_\xi\eta-\xi\beta_\xi^2)+
\beta_\eta\beta_\xi-\beta_\eta\beta_{\xi,x}+\beta_{\eta,x}\beta_\xi=F_2(\la),
\end{split}
\end{equation}
for some suitable functions $F_1(\la), F_2(\la)$, independent of
$x$. It is not difficult to ascertain that the choice $F_1(\la)=0,
F_2(\la)=-\rho\la$ is a good one. Indeed one can show that, with
such a choice, we have
\begin{enumerate}
\item $\beta_\eta^0=\fddd{H_{-1}}{\eta},\>
\beta_\xi^0=\fddd{H_{-1}}{\xi}$, with
$H_{-1}=\int\sqrt{\eta^2+4\rho\xi}\, dx$. \item The coefficients
$\beta_\eta^i,\beta_\xi^i$ can be algebraically found
  from eq. \rref{eq:casint} (with  $F_1(\la)=0, F_2(\la)=-\rho\la$) as differential
  polynomials of $\{\beta_\eta^j,\beta_\xi^j\}_{j=0,\ldots,i-1}$.
\end{enumerate}
Actually, more is true. Indeed it holds:
\begin{prop}
The normalized equations
\begin{equation}
  \label{eq:ncasint}
  \begin{split}
&2\rho\la\beta_\eta+\beta_\xi(1-\la\eta)-\beta_{\xi,x}=0\\
&\la(\rho\beta_\eta^2-\beta_\eta\beta_\xi\eta-\xi\beta_\xi^2)+
\beta_\eta\beta_\xi-\beta_\eta\beta_{\xi,x}+\beta_{\eta,x}\beta_\xi+\rho\la=0
\end{split}
\end{equation}
give the recursive equations for the Casimir function of the Poisson
pencil $P_\la$, that has the form
\begin{equation}
  \label{eq:hcas}
  \CH_-(\la)= \int \frac{2\rho}{\beta_\xi}\, dx.
\end{equation}
\end{prop}
{\bf Proof}. Solving for $(\eta,\xi)$ the equations
\rref{eq:ncasint}, we get, in terms of $\beta_\xi$ and
$h=\beta_\eta/\beta_\xi$ the equations
\begin{equation}\label{ex-bh}
\begin{split}
&\eta=2\rho h + \frac{1}{\la} \big(1-(\log{\beta_\xi})_x\big)\\
&\xi= -\rho\big( {h}^2-\frac{1}{\beta_{\xi}^{2}}\big)+
\frac{1}{\la} \big( h_x+ h(\log{\beta_\xi})_x\big)
\end{split}
\end{equation}
So, if $(\dot \eta, \dot \xi)$ is any tangent vector to the
manifold of pairs $(\eta,\xi)$ we get
\begin{equation}\label{eq:ui}
\begin{split}
&\dot{\eta}=2\rho \dot{h} +
\frac{1}{\la}\ddt{}((\log{\beta_\xi})_x)
\\&\dot{\xi}=
-\rho\big(2h\dot{h}+\frac{2\dot{\beta}_\xi}{\beta_{\xi}^{3}}\big)+
\frac{1}{\la} \big( \dot{h_x}+ \dot{h}(\log{\beta_\xi})_x
+h\ddt{}((\log{\beta_\xi})_x) \big).
\end{split}
\end{equation}
A straightforward computation shows that
\begin{equation}\label{endcomp}
\dot\eta\beta_\eta+\dot\xi\beta_\xi\equiv \dot\eta
h\beta_\xi+\dot\xi\beta_\xi= -2\rho
\frac{\dot{\beta_\xi}}{\beta_{\xi}^2}+\frac{\big(\dot{h}b\big)_x}{\la}
\end{equation}
whence the assertion.

The first coefficients of the Casimir $H(\la)$ of the negative
\ger y beyond $H_1=\int\sqrt{4\rho\xi+\eta^2}$ can be explicitly
computed  as
\begin{equation}\begin{split}
&H_{-2}=-\int \frac{\eta+\eta_x}{\sqrt{4\rho\xi+\eta^2}},\\
&H_{-3}=
\int\left(\frac{(2\rho(\xi+\xi_{xx})+\frac12\eta_x^2-\eta(\eta_x-\eta_{xx}))}{(4\rho\xi+\eta^2)^{3/2}}-\frac52\frac{(2\rho\xi_x+\eta\eta_x)^2}{(4\rho\xi+\eta^2)^{5/2}}
\right),\end{split}  \end{equation}
the expression of the subsequent coefficients being
too complicated to be usefully reported.
\subsection{Lax representation}\label{subsec4.1} The GZ analysis
of the Poisson pencil $P_\la$ of \rref{eq:mapenc} provides a way
to find a Lax representation for the CH2 equations
\rref{eq:zlfch-eq}. Indeed, as we are considering a Casimir of
$Q-\la P$, the "first integrals" corresponding to \rref{eq:casint}
acquire the form:
\begin{equation}\label{eq:n1}
    \left\{
    \begin{array}{l}
2\rho\beta_\eta+\beta_\xi(\la-\eta)-\la\beta_{\xi,x}=f_1(\la)\\
\rho\beta_\eta^2-\beta_\eta\beta_\xi\eta-\xi\beta_\xi^2+\la(
\beta_\eta\beta_\xi-\beta_\eta\beta_{\xi,x}+\beta_{\eta,x}\beta_\xi)=f_2(\la).
    \end{array}
    \right.
\end{equation}
Since the Casimir of the pencil starts with $\int \xi \, dx$, a
consistent normalization is $f_1(\la)=\la$, $f_2(\la)=0$. Let us
define
\begin{equation}\label{eq:hdef}
    h=\beta_\eta/\beta_\xi.
\end{equation}
Dividing the second of equations~\rref{eq:n1} by $({\beta_\xi})^2$
we can rewrite it as a Riccati equation for $h$:
\begin{equation}\label{eq:rh}
    \rho h^2-\eta h -\xi+\la(h_x+h)=0,
\end{equation}
where the right hand side can be set to zero consistently
with the normalization of $f_1(\la), f_2(\la)$.
In turn, setting
\begin{equation}\label{eq:hpsi}
\psi=\exp \frac{\rho}{\la}\int^x h,
\end{equation}
equation \rref{eq:rh} linearizes as
\begin{equation}\label{eq:l}
    \psi_{{{xx}}}= \left( \dsl{\frac {\eta}{\lambda}}-1 \right) \psi_{{x}}
+\dsl{{\frac {\xi\,\rho}{{\lambda}^{2}}}} \psi.
\end{equation}
This scalar operator is the first member of the Lax pair for CH2.

To get the second member, we substitute \rref{eq:hpsi} into
\rref{eq:hdef}, and notice that the differential of $H^+(\la)$
starts with
\begin{equation}\label{eq:n2}
    (\frac{u}{\la}, 1+\frac{v}{\la}).
\end{equation}
Since CH2 can be, using standard techniques of the \bih\ theory,
written as a Hamiltonian equation {w.r.t. the Poisson pencil}
$Q-\la P$ with $\la$--dependent Hamilton function
\begin{equation}\label{eq:n3}
    \CH(\la)=\int(\la u+u\eta) dx,
\end{equation}
we arrive at the second member of the Lax pair for CH2, given by
the linear equation
\begin{equation}\label{eq:lt}
    \psi_t=\left( \lambda+v \right) \psi_{{x}}-{\frac
    {u\rho}{\lambda}}\psi,
\end{equation}
interpreting Eq. \rref{eq:hdef} as
\[
\fddd{\CH}{\xi}h-\fddd{\CH}{\eta}\sim \del_t\log{\psi}.
\]
\begin{prop}\label{prop:lax}
The CH2 equations
\[
(v- v_{x})_{t}={(-2\rho
    u+v^2- vv_x)_x};\quad
    {(u+ u_{x})_{t}}=(2uv+ u_x v)_x
\]
are the compatibility conditions of the two linear equations:
\begin{equation}\label{eq:cc}
\psi_{{{xx}}}= \left( \dsl{\frac {\eta}{\lambda}}-1 \right)
\psi_{{x}} +\dsl{{\frac {\xi\,\rho}{{\lambda}^{2}}}} \psi;\quad
\psi_t=\left( \lambda+v \right) \psi_{{x}}-{\frac
    {u\rho}{\lambda}}\psi.
\end{equation}
\end{prop}
For further reference, as well as for the reader's convenience, we
remark that CH2 can be seen as the compatibility condition for the
two first order matrix linear operator $\del_x-L$ and $\del_t-V$,
where
\begin{equation}\label{eq:lax2}
L= \left( \begin {array}{cc} 0&1\\\noalign{\medskip}\dsl{{\frac
{\rho\,\xi}{{\lambda}^{2}}}}&\dsl{{\frac {\eta}{\lambda}}}-1\end
{array} \right), \quad V= \left( \begin {array}{cc} -\dsl{\frac
{\rho\,u}{\lambda}}&\lambda+v
\\\noalign{\medskip}\rho\, \left( \dsl{\frac {u}{\lambda}}+\dsl{\frac {v\xi}{{
\lambda}^{2}}} \right) &-\lambda+\dsl{\frac
{v\eta}{\lambda}}-\dsl{\frac {\rho \,u}{\lambda}}\end {array}
\right).
\end{equation}

\section{Solutions}\label{sec:5}
The aim of this Section is to describe a few solutions of the CH2
equations. We start with cnoidal waves. Searching for solutions of
the form
\begin{equation}\label{5.1}
u=u(x-ct),\> v=v(x-ct)
\end{equation}
we have
\begin{equation}\label{5.2q}
\begin{array}{l}{c(v'-v)=(-2\rho u+v^2-v'v)+\gamma_1}\\{c(u+u')=-(2vu+u'v)-\gamma_2},
\end{array}\Rightarrow \begin{array}{l}
v'=\dsl{\frac{-2\rho u+v^2+cv+\gamma_1}{(c+v)}}\\
u'=-\dsl{\frac{2vu+cu+\gamma_2}{c+v}}.
\end{array}
\end{equation}
Multiplying the first of \rref{5.2q} by $u'$ and the second
by $v'$ and subtracting the two equations we get
\begin{equation}\label{5.3}
-c(uv'+u'v)=-2\rho u'u+u'v^2+2vv'u+\gamma_1u'+\gamma_2 v',
\end{equation}
that integrates to the elliptic curve $\CE$
\begin{equation}\label{5.3a}
cuv -\rho u^2+uv^2+\gamma_1 u+\gamma_2 v+\gamma_3=0.
\end{equation}
It is easy to ascertain that \rref{5.2q} expresses the cnoidal wave
solutions of CH2 via third kind Abelian integrals on $\CE$,
as in the Camassa-Holm case.

\subsection{The traveling wave solution}\label{ssec:peak}
We can find the traveling wave solution of the CH2 equations
setting to zero the three constants $\gamma_i, i={1,2,3}$ of Eq. \rref{5.3}, 
that is, considering the degenerate curve $ u(cv -\rho
u+v^2)=0.$
Let us suppose $c>0$ and $\rho>0$.\\
A solution for $u=u(r)$ is
\begin{equation}\label{eq:su}
u(r)=  \left( -\frac{c}{\sqrt {\rho}}+\kappa\,{e^{-r}} \right)
\kappa\,{e^{-r }},
\end{equation}
where $\kappa$ is a (positive) constant.

This solution vanishes for
for $r\to+\infty$ and for
\[
r=r_0=-\log{\dsl{\frac{c}{\kappa\sqrt{\rho}}}}.
\]
It is negative for $r> r_0$ and
has an absolute minimum for
\[
r=r_{min}=-\log{\dsl{\frac{c}{2\kappa\sqrt{\rho}}}}.
\]
This means that a bounded continuous solution for the traveling
wave of CH2 can be obtained joining the trivial solution
\[
u=0 \quad \text{for  }\>  r\le r_0
\]
with this one, that is, considering
\begin{equation}\label{eq:ur}
u(r)=  \left( -\frac{c}{\sqrt {\rho}}+\kappa\,{e^{-r}} \right)
\kappa\,{e^{-r }}\theta(r-r_0),
\end{equation}
where $\theta(r)$ is the Heaviside step function. Substituting
this result in the equations \rref{5.2q} we see that we can find a
bounded continuous solution for $v(r)$ as well in the form of a
ordinary (although well-shaped) peakon
\begin{figure}
\caption{CH2 peakon profiles with $c=1,\rho=0.1,\> 0.5,\> 1$. The
thick line is $u(s)$, the thin line is $v(s)$} \vspace{0.8truecm}
\boxed{\centerline{\epsfxsize=4.8cm\epsfbox{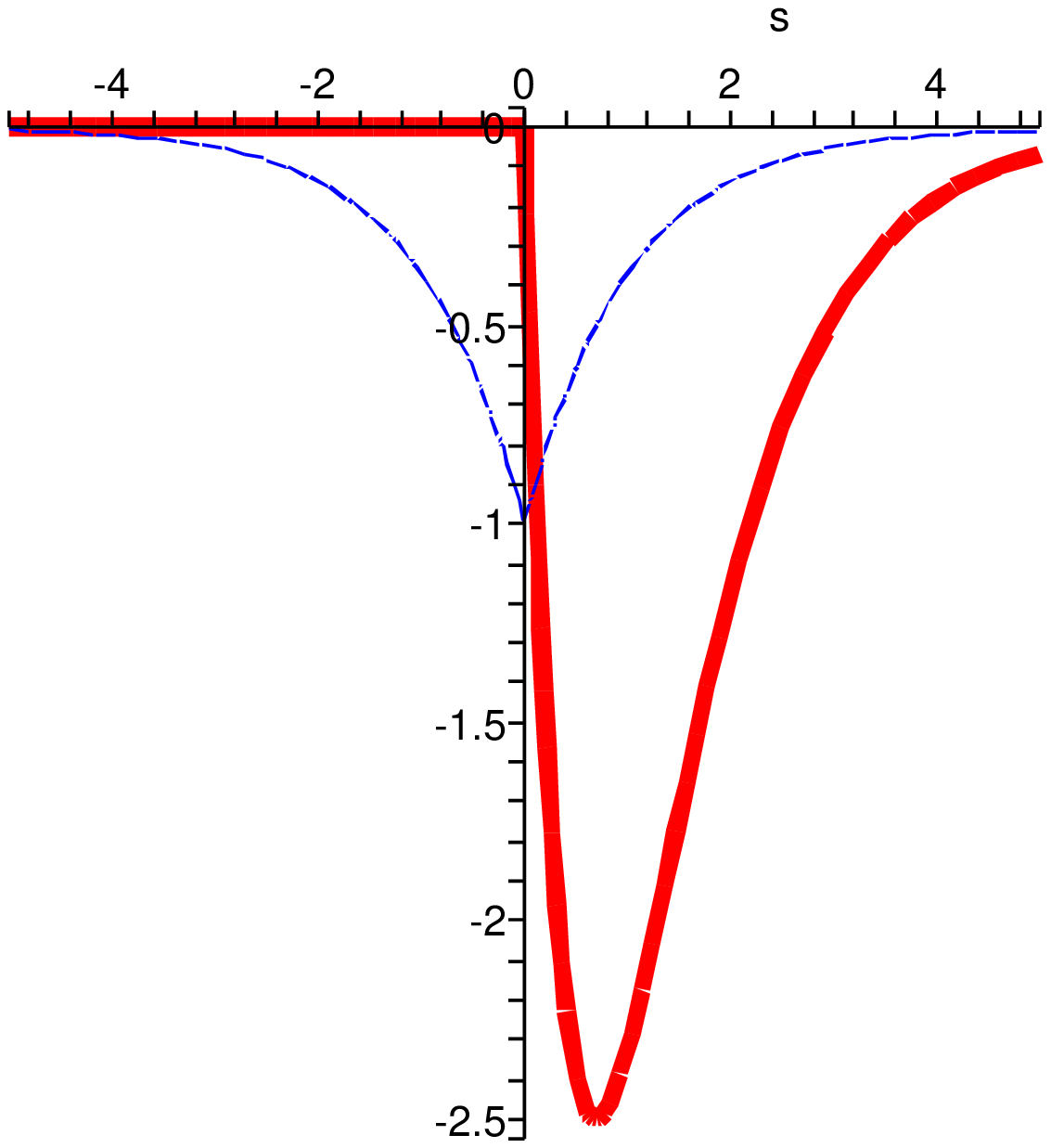},\quad
\epsfxsize=4.8cm\epsfbox{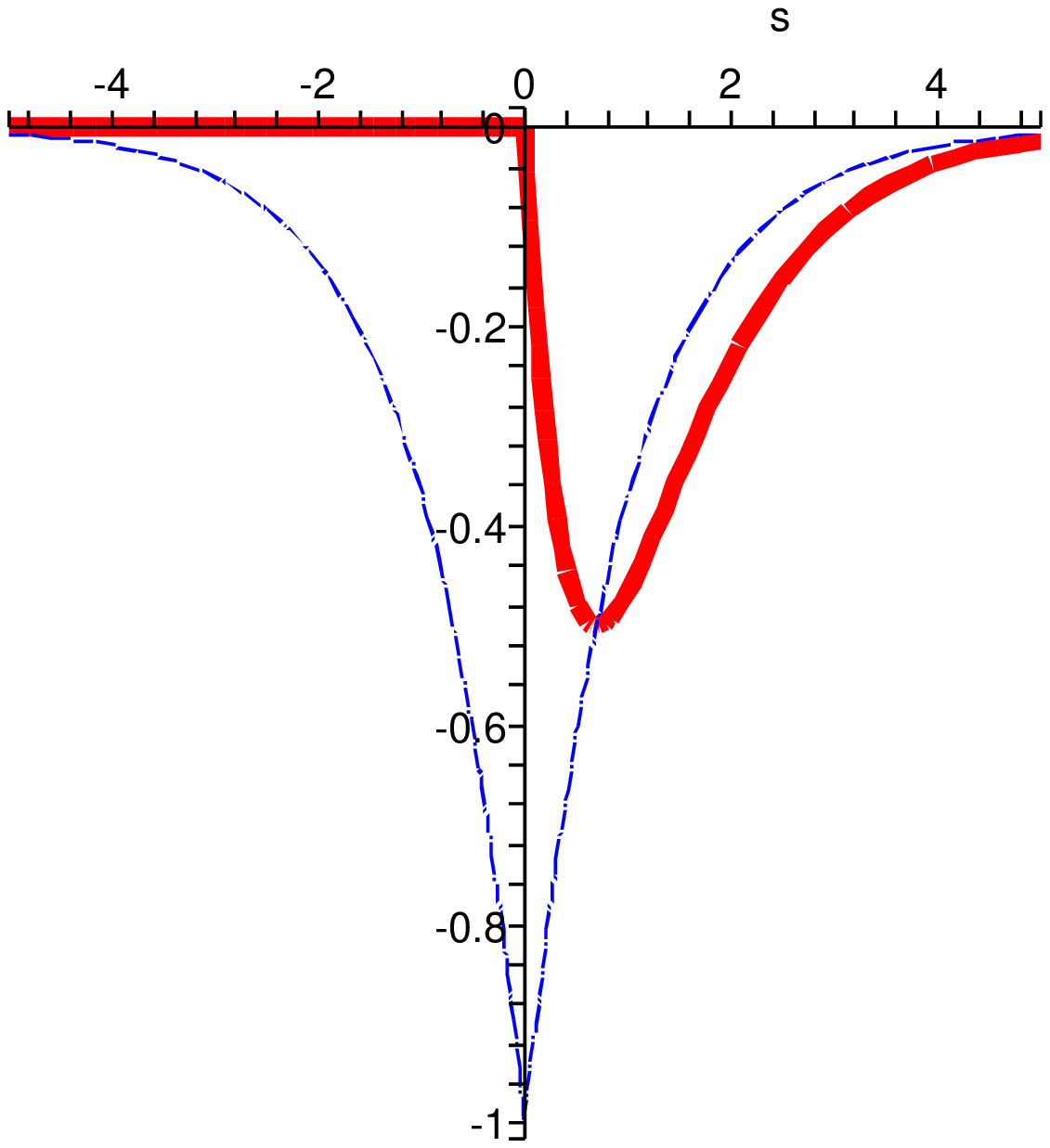}\quad
\epsfxsize=4.8cm\epsfbox{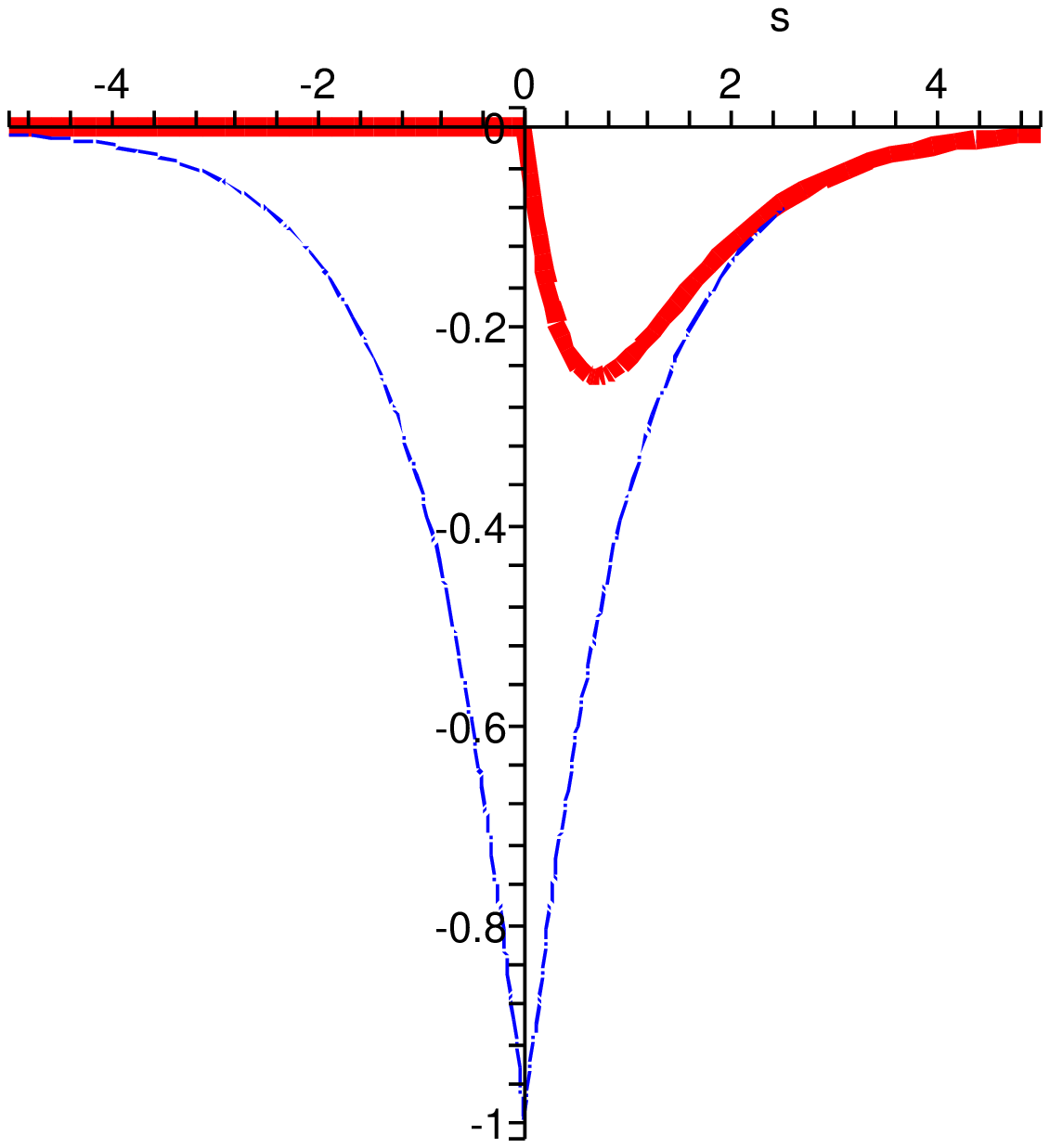}}}
\end{figure}
\begin{equation}\label{eq:vr}
\left\{\begin{array}{l}
v(r)=-\dsl{\frac{c^2}{\kappa\sqrt{\rho}}}\expo{r},\quad r\le r_0\\
v(r)=-\kappa\sqrt{\rho}\expo{(-r)},\qquad r\ge
r_0,\end{array}\right.
\end{equation}
In terms of the normalized variable $s=r-r_0$, we can compactly
write this peakon solution as
\begin{equation}\label{eq:rpeak}
u(s)=\dsl{\frac{c^2}{\rho}}(e^{-2s}-e^{-s})\theta(s),\> v(s)=-c
e^{-|s|}, \quad s=x-ct+r_0.
\end{equation}
We notice that the derivatives of both $u(s)$ and $v(s)$ have a
discontinuity in $s=0$ and that Rankine-Hugoniot type conditions
are satisfied at $s=0$. The profile of three such peaked solutions
are depicted in Figure 1.

Mimicking the Camassa-Holm original paper, we can seek for
$N$-peakon solutions of CH2 by means of a linear superposition of
elementary solutions of the form given by Eq. \rref{eq:rpeak},
.i.e., via the {\em Ansatz}
\begin{equation}\label{eq:Npe}
\begin{array}{l}
u(x,t)=\sum_{i=1}^{N}
\pi_i(t)(\expo{(-2x+q_i(t))}-e^{(-x+q_i(t))})\theta(x-q_i(t)),\\
v(x,t)=\sum_{i=1}^N -p_i(t)e^{(-|x-q_i(t)|)}.
\end{array}
\end{equation}
A tedious but straightforward computation shows the following:
\begin{prop}\label{prop:N-peak}
The Ansatz \rref{eq:Npe} gives solutions of the CH2 equation
provided that the $2N$ dynamical variables $(q_i, p_i)$ evolve as
\begin{equation}\label{eq:pq}
\dot{q_i}=\sum_j p_j\expo{(-|q_i-q_j|)};\quad  \dot{p}_i=p_i\sum_{j=1}^N
p_j\expo{(-|q_i-q_j|)}\wid{\epsilon}_0(q_i-q_j),
\end{equation}
that is, they evolve, as in the CH case \cite{CH93} as canonical
variables under the Hamiltonian flow of
\[
H_N=\frac12 \sum_{i,j=1}^N p_ip_j\expo{(-|q_i-q_j|)}.
\]
The quantities $\pi_i$ can be obtained from the solutions
$(q_i(t),p_i(t))_{i=1\ldots N}$ of \rref{eq:pq} via the formula:
\begin{equation}\label{eq:pi}
\pi_i=\dsl{\frac{2p_i}{\rho}}\sum_{j=1}^N
p_j\expo{(q_j-q_i)}\wid{\theta}_{\frac12}(q_i-q_j).
\end{equation}
In these formulas, $\wid{\theta}_{\frac12}(s)$ is the Heaviside step
function with the normalization $\wid{\theta}_{\frac12}(0)=\frac12$,
and
$\wid{\epsilon}_0(s)=\wid{\theta}_{\frac12}(s)-\wid{\theta}_{\frac12}(-s)$.
\end{prop}
\noindent {\bf Remark}. The coincidence of the peakon solution of
CH2 with those of CH can be, {\em a posteriori},  understood by
means of the following direct link between CH2 and CH, which we
report from \cite{LZ04}. If we apply the operator
$\mathbf{1}+\del_x$ to the first of Eq \rref{eq:zlfch-eq}, we get
\[\begin{split}
&(v-v_{xx})_{t}=\del_x(-2\rho(u+ u_x)+
vv_x+v^2-(vv_{xx} + v_x^2))\\
&=\del_x(w+(\frac32v^2-(\frac12 v_x^2+vv_{xx}))).\end{split}
\]
Since the additional term $w$ in this equation is given by
\[
w=-2\rho(u+ u_x)-\frac12(v- v_x)^2
\]
we deduce the following system:
\begin{equation}
  \label{eq:zch}
  \begin{split}
&(v-v_{xx})_{t}=\del_x(w+(\frac32v^2-(\frac12
v_x^2+vv_{xx})))\\
&w_{t}=v_{t}w+2w_{t}v,\end{split}
\end{equation}
whose consistent reduction to $w=0$ exactly gives the CH
equation~\rref{eq:che}.

\subsection{Finite gap solutions}
In this Section we will address the basic properties of ``finite gap''
solutions of the CH2 \ger y.

To this end, we use the Lax representation of the \ger y, i.e.,
the vanishing curvature relations (see the end of Subsection
\ref{subsec4.1})
\begin{equation}\label{eq:0cur}
\del_t L-\del_x V+[L,V]=0, \quad\text{with } L=\left( \begin
{array}{cc} 0&1\\\noalign{\medskip}\dsl{{\frac
{\rho\,\xi}{{\lambda}^{2}}}}&\dsl{{\frac {\eta}{\lambda}}}-1\end
{array} \right),
\end{equation}
and $V$ given in Eq. \rref{eq:lax2}.

Following standard techniques in the theory of integrable
systems, we consider a matrix $2\times 2$ matrix $W(\la)$, with
elements that depend on $\la$, such that the Zakharov-Shabat
relations
\begin{equation}\label{eq:zash}
W(\la)_x-[L({\la}),W(\la)]=\frac{\del}{\del t_W} L(\la)
\end{equation}
yield consistent equations for the variables $\eta,\xi$.

A straightforward computation shows
\begin{prop}\label{prop:0c}
To yield consistent Zakharov-Shabat equations~\rref{eq:zash} the
matrix $W(\la)$ must have the form
\begin{equation}\label{eq:wmat}
W(\la)= \left( \begin {array}{cc} -\dsl{{\frac {\rho\,a  }{\lambda
}}}&b
\\\noalign{\medskip}-\dsl{{\frac{\rho {a_x  }}{\lambda}}}+\dsl{{\frac {\rho\,\xi  b
  }{{\lambda}^{2}}}}&{b_x
-b } +\dsl{{\frac {b  \eta -\rho\,a }{\lambda}}}\end {array}
\right),
\end{equation}
with the "time" evolution $\ddd{}{t_W}$ given by
\begin{equation}\label{eq:gh}
\vec2{\ddd{\eta}{t_W}}{\ddd{\xi}{t_W}}=\left(Q-\la
P\right)\vec2{a}{b}
\end{equation}
This means that $a=a(x,\la)$ and $b(x,\la)$ are obtained form the
Casimir of the Poisson pencil $\CH(\la)=\CH_0+\frac1\la
\CH_1+\cdots $ via:
\[
a(x,\la)=[p(\la)(\fddd{\CH(\la)}{\eta})]_+, \quad
b(x,\la)=[p(\la)(\fddd{\CH(\la)}{\xi})]_+,
\]
for some polynomial (with constant coefficients) $p(\la)$, where
$[f(\la)]_+$ represents the
polynomial part in the expansion in $\la$ of
$f(\la)$.
\end{prop}
{\bf Remarks. 1)} The Zakharov-Shabat representation for the times
of the negative \ger y of Subsection \ref{subs:3.1} can be obtained in a similar manner.
\\ 
{\bf 2)} Eq. \rref{eq:n1} with the normalization $f_1(\la)=\la,
f_2(\la)=0$, reads $\text{Tr}(W(\la))=1$. Solving this  for
$a(x,\la)$, we can express every element in terms of $b(x,\la)$ as
follows:
\begin{equation} \label{eq:nw}
\begin{split} & W(\la)=\frac12
 \left( \begin {array}{ccc} -b(x,\la)_{x}+b(x,\la)(1-\dsl{{\frac{\eta}{
\lambda}}})-1&&2b(x,\la)\\
W_{2,1}&& W_{2,2}
\end {array} \right),\\&
W_{2,1}=b(x,\la)_{x}-\,b(x,\la)_{x\,
x}-(b(x,\la)\eta)_x/\la+(2\rho\,\xi\,b(x,\la))/\la^2,\\&
W_{2,2}=b(x,\la)_{x}-b(x,\la)(1-\eta /\lambda)-1.
\end{split}
\end{equation}
{\bf 3)} In the notations of Section \ref{sec:2}, the
Lax representation of the $n$-th flow of the \ger y\footnote{We
use the convention that
  $\ddd{}{t_1}=\ddd{}{x}$, $\ddd{}{t_2}=CH2$.} will be obtained choosing
\begin{equation}\label{eq:ab}
a_n(x,\la)=[\la^{n-1}\beta_\eta]_+,  b_n(x,\la)=[\la^{n-1}\beta_{\xi}]_+.
\end{equation}
We notice that, in this way, $b_n(x,\la)$ is a monic polynomial of
degree $n-1$ (while $a_n(x,\la)$ is of  $n-2$). For further
use we recall that, if $n\ge k$, the following relation holds, as
a consequence of the definitions:
\begin{equation}\label{eq:bnk}
b_k(x,\la)=\Big[\dsl{\frac{b_n(x;\la)}{\la^{n-k}}}\Big]_+.
\end{equation}

As it has been already noticed, the recurrence relations for the
differentials of the Casimir function of the {\em positive \ger y}
-- that is, the \ger y to which the CH2 equations belong - cannot
be solved, via local functionals in $u,v$, let alone in
$\xi,\eta$. However, we can still use the Zakharov-Shabat
representation of the \ger y in the study of (some properties of)
the {\em stationary} $t_n$ submanifolds for CH2.

Indeed, applying a scheme that dates back to \cite{Al79} and has
been used in \cite{achm94,acfhm01} for studying finite-gap
solutions of Harry-Dym and Camassa-Holm type, we can argue as
follows.

We consider $L=W_1$ and the matrices $W_n$ associated with the
time $t_n$ of the \ger y\footnote{From now on we will not
explicitly write the dependence on $(x, \la)$ of the polynomial
$b$ anymore.}. On the stationary manifold of the time $t_n$ the
matrix
\begin{equation}\label{eq:Wn}
W_n=\frac12\mat2{b_n(1-\eta/\la)-b_{nx}-1}{2b_n}{b_{n x}-b_{n\,
xx}-(b_n\eta)_x/\la +2\rho\xi/\la^2}{b_n(\eta/\la-1)+b_{nx}-1},
\end{equation}
where $b_n=[\la^{n-1}\beta_\xi]_+$,
undergoes, along the time $t_m$ of the \ger y, the Lax equation
\begin{equation}\label{eq:laxwn}
\ddd{}{t_m}{W_n}=[W_m,W_n],
\end{equation}
and so its spectral curve $\Gamma=\text{det}(\nu-W_n(\la))$,
is a constant of the motion. Defining
$\nu=\dsl{\frac{\mu-\la}{2\la}}$, $\Gamma$ is
given by $\mu^2=R(\la)$, where
\begin{equation}\label{eq:spcu}
R(\la)=(b_n^2-2b_nb_{n\,xx}+
b_{n\,x}^2)\la^2-2b_n^2\big((\eta-\eta_x)\la-(4\rho\xi+\eta^2)\big).
\end{equation}
 We notice that, since $b_n$ is a monic
polynomial of degree $n-1$ in $\la$, the degree of $R(\la)$ is $2n$,
and so the genus of $\Gamma$ is $n-1$.

Let $(\la_1,\ldots,\la_{n-1})$ be the $n-1$ roots of $b_n$, i.e, $
b_n=\dsl{\prod_{j=1}^{n-1}(\la-\la_j)}$.
The $x=t_1$ evolution equation of these roots is governed
by the equation
\begin{equation}\label{eq:xev}
\la'_i=\ddd{\la_i}{x}=\dsl{\frac{\sqrt{(R(\la_i))}}{\la_i\prod_{j\neq
i}^n(\la_i-\la_j)}}
\end{equation}
Moreover, thanks to the form of the matrices $W_n$, the Lax
equations~\rref{eq:laxwn} imply that $b_n$ evolves along $t_k$ as
follows
\begin{equation}\label{eq:bntk}
\ddd{b_n}{t_k}=b_k\ddd{b_n}{x}-b_n\ddd{b_k}{x}, \qquad k=1,\ldots, n-1,
\end{equation}
and so we get, evaluating these relations for $\la=\la_i$,
\begin{equation}\label{eq:btk}
\ddd{\la_i}{t_k}=b_k(\la_i)
\ddd{\la_i}{x}= b_k(\la_i)
\dsl{\frac{\sqrt{(R(\la_i))}}{\la_i\prod_{j\neq
i}^n(\la_i-\la_j)}}.
\end{equation}
Since the equations~\rref{eq:bnk} imply that $b_k(\la_i)$ is the
symmetric polynomial of order $k-1$ in
$\big(\la_1,\ldots,\la_{i-1},\la_{i+1},\ldots,\la_{n-1}\big)$
(see, e.g., \cite{acfhm01} for the details), one arrives at the
following relations
\begin{equation}\label{eq:nsjacobi}
dt_k=\sum_{i=1}^{n-1}
\dsl{\frac{\la_i^{n-k}d\la_i}{\sqrt{R(\la_i)}}}.
\end{equation}
These are non-standard Abel-Jacobi relations on the genus $g=n-1$
curve $\Gamma(\mu,\la)$, entailing that  the third kind Abelian
differential $\frac{\la^g\,d\la}{\sqrt{R(\la)}}$ is involved in
the integration of the finite dimensional dynamical system
associated with the stationary $t_n$ manifold of CH2.

\subsection{The manifold $\mathbf{u=0}$ and singular solutions}
The CH2 equations admit a consistent reduction to the manifold
$u=0$. In this section we will show that the resulting equation,
namely
\begin{equation}\label{eq:gf}
\partial_t(v-v_x)=\partial_x(v^2-v v_x)
\end{equation}
admit, with a suitable interpretation, non-continuous solutions,
that could be called half-peakons or {\em "cliffon"} solutions\footnote{We borrow the name
  from the paper \cite{DHH02}.}. We notice that the opportunity
of studying this equation (or rather the variant of it therein
considered) was suggested in \cite{OR95}.

We start noticing that the Green operator for $\mathbf{1}-\del_x$
is the following:
\begin{equation}\label{eq:green}
G(f)(x)=\int_x^\infty\expo{x-y}f(y) dy=\int_{-\infty}^\infty
\expo{x-y}\theta(y-x) f(y) dy,
\end{equation}
where $\theta(r)$ is the Heaviside step function.

By means of this Green operator, we can rewrite Eq. \rref{eq:gf}
in the following integral form:
\begin{equation}\label{eq:gf-i}
\partial_t v=\partial_x\big(\frac{v^2}{2}+G(\frac{v^2}2)\big),
\end{equation}
to be considered as the analogue of Eq. (0.1) of \cite{CMc99}.
With the Ansatz
\begin{equation}\label{eq:an}
v(x,t)=A\expo{(x-ct)}\theta(ct-x),
\end{equation}
we get
\begin{equation}\begin{split}
&G(v^2)(x)=A^2\,\int_{-\infty}^\infty
\expo{(x+y-2ct)}\theta(y-x)\theta(ct-y) dy=\\
&
A^2\theta(ct-x)\big(\expo{(x-ct)}-\expo{(2x-2ct)}\big)=-v^2+Av.\end{split}
\end{equation}
so that bounded traveling waves for Eq.
\rref{eq:gf} are of the form
\begin{equation}
v(x,t)=-2c \expo{(x-ct)}\theta(ct-x),
\end{equation}
that is, well-shaped peakons (with amplitude twice their speed),
for $x< ct$ that drops to zero for $x> ct$. One can notice the
same solution can be gotten from the differential
equation\rref{eq:gf}, provided we agree to regularize the product
of an Heaviside function and a Dirac $\delta$-function as
$\theta(r)\delta(r)=\frac12 \delta(r)$.

The interaction of $N$ right moving cliffons can be described much
in the same way of that concerning $N$ right moving peakons,
described in Subsection \ref{ssec:peak}, via the Ansatz:
\begin{equation}\label{a:N}
v(x,t)=\sum_{i=1}^N -2p_i(t)\expo{(x-q_i(t))}\theta(q_i(t)-x).
\end{equation}
One should notice the following. To evaluate $G(v^2)$ we have to
consider products
\begin{equation}
F_{ij}=p_ip_j\expo{(2x-q_i-q_j)}\theta(q_i-x)\theta(q_j-x).
\end{equation}
They can be written as
\[
F_{ij}=p_ip_j\expo{(2x-q_i-q_j)}\big(\wid{\theta}_{\frac12}(q_i-q_j)\theta(q_j-x)+
\wid{\theta}_{\frac12}(q_j-q_i)\theta(q_i-x)\big),
\]
where the normalized Heaviside function
$\wid{\theta}_{\frac12}(r)$ has been defined in Proposition
\ref{prop:N-peak}.
Since a straightforward computation yields
\begin{equation}
\label{eq:boh}\begin{split} G(F_{ij})(x)+F_{ij}&(x)=\\
p_ip_j\big(\expo{(x-q_j)}\theta(q_i-x)\wid{\theta}_{\frac12}(q_j-q_i)&+
\expo{(x-q_i)}\theta(q_j-x)\wid{\theta}_{\frac12}(q_i-q_j)\big),\end{split}
\end{equation}
we obtain that the resulting dynamical system for the $2N$
quantities $(q_i,p_i)$ is the following {\em impulsive} system:
\begin{equation}\label{eq:dynsyst}
\left\{\begin{array}{l} \dot{p_i}=0\\
 \dot{q_i}=2\sum_j
p_j\expo{q_i-q_j}\wid{\theta}_{\frac12}(q_j-q_i).\end{array}\right.
\end{equation}
\begin{figure}
\caption{2-cliffon interaction: thin lines represent the
individual cliffons, while the thick one their superposition.}
\vspace{1.truecm}\boxed{
\centerline{\epsfxsize=4.8cm\epsfbox{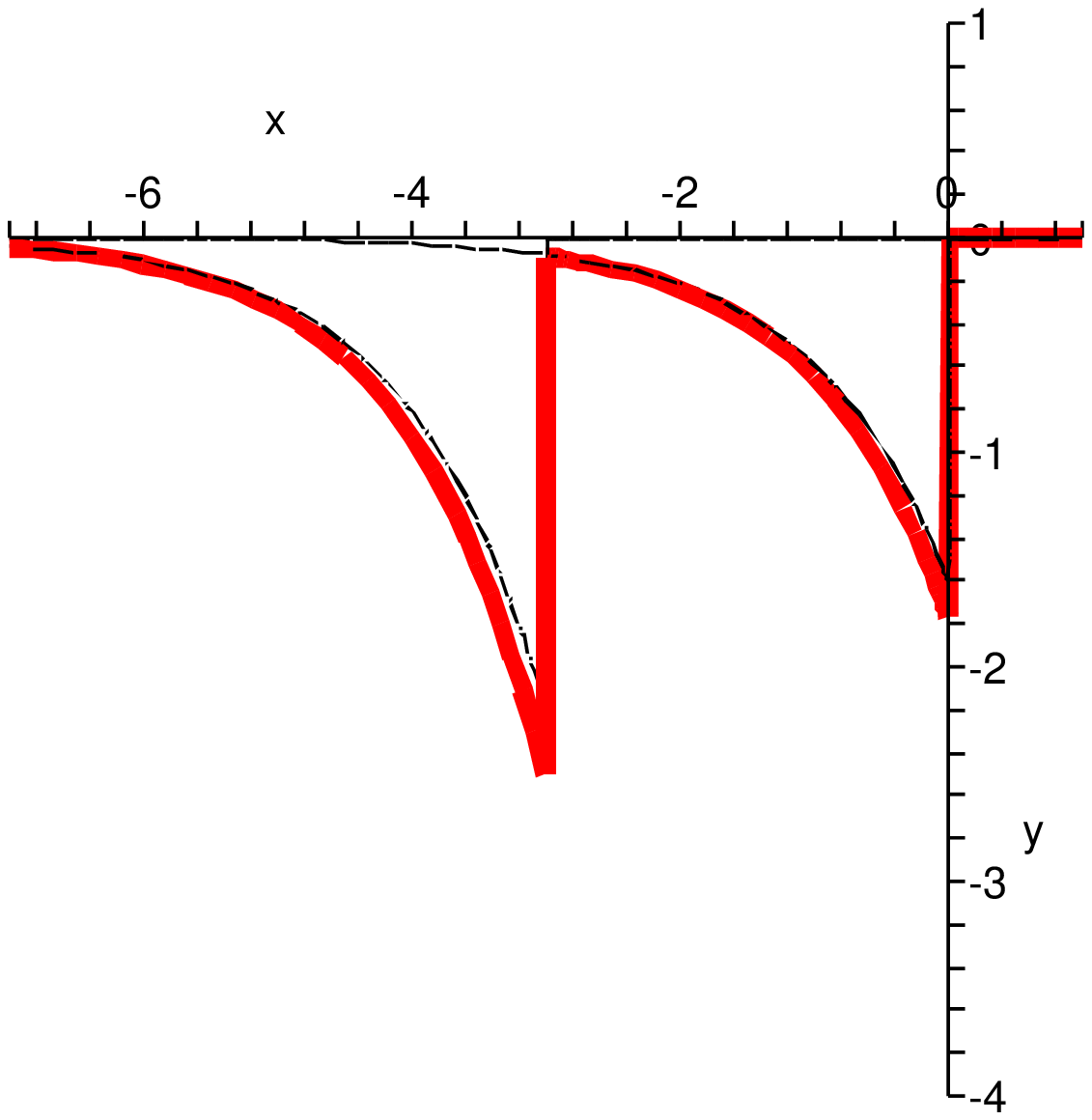},\quad
\epsfxsize=4.8cm\epsfbox{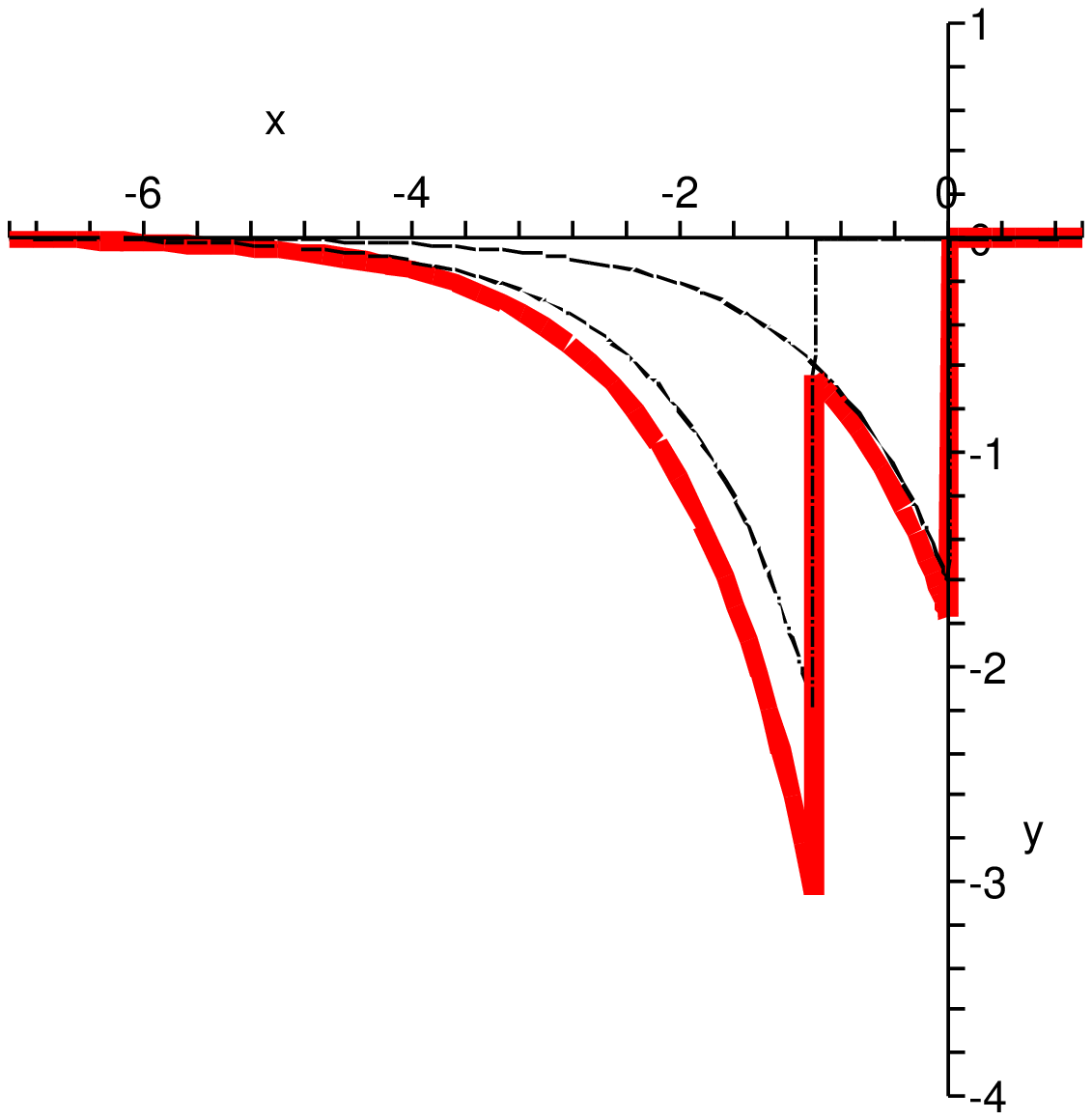}\quad
\epsfxsize=4.8cm\epsfbox{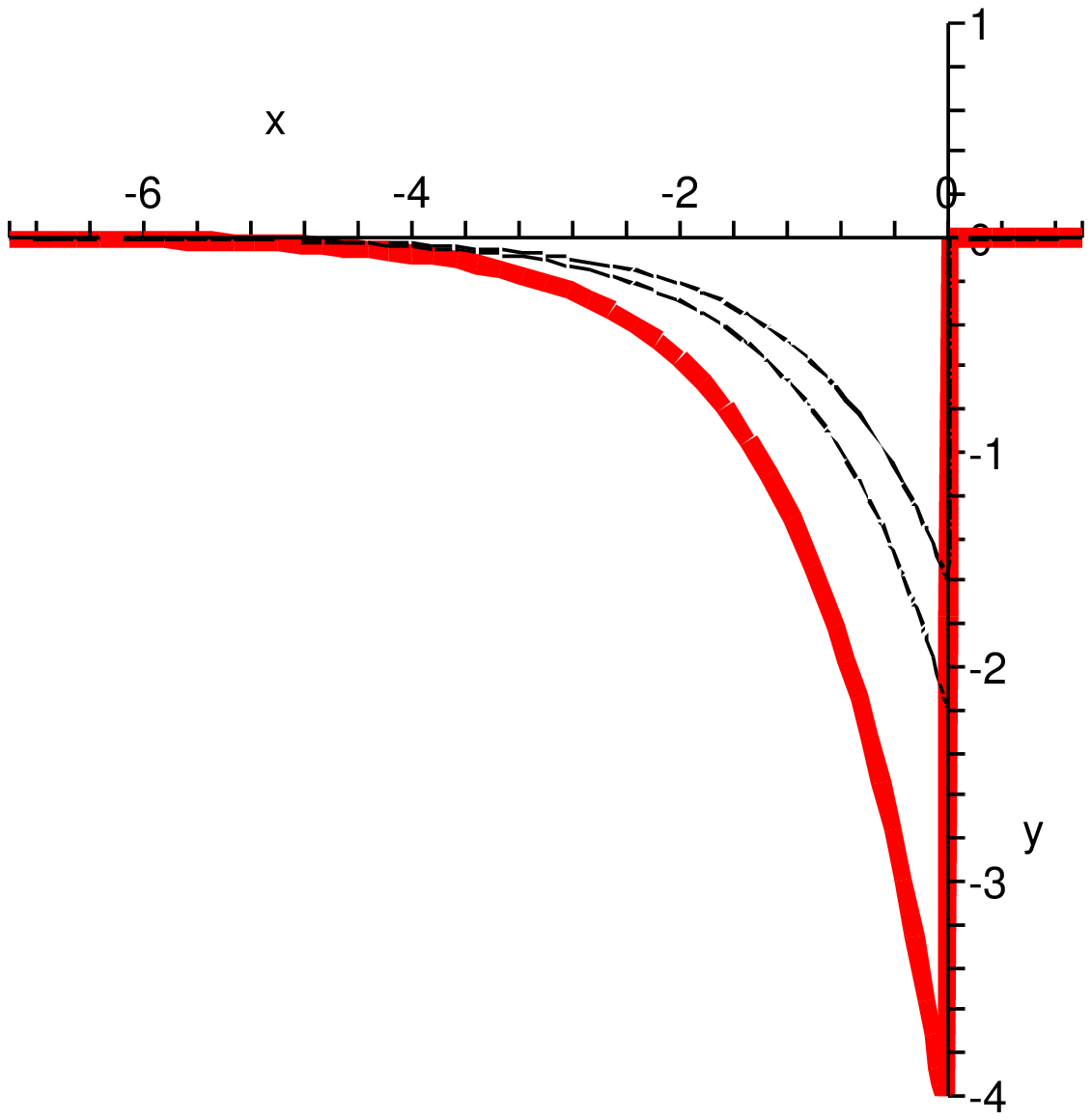}}}
\end{figure}
We close the paper examining these equations of motion for a
system of two cliffons, with initial conditions
\[
q_1(0)\ll q_2(0), p_1(0)\simeq p_2(0)
\]
For $t\simeq 0$ the evolution will be as follows (notice that
$p_i(t)\equiv p_i(0)$):
\[
\dot{q}_1=p_1+2p_2\expo{q_1-q_2}, \quad \dot{q}_2=p_2.
\]
This means that the rightmost cliffon (whose tip is located at
$x=q_2(t)$) will travel undisturbed, while the leftmost one is
accelerated towards the rightmost. The initial acceleration is
small, but positive. When the distance $q_1-q_2$ between the two
tips is small but non zero, the acceleration of the leftmost
cliffon will become sensible, so that the asymptotic equation is
\[
\dot{q}_1\simeq p_1+2p_2, \quad \dot{q}_2=p_2,\quad\text{for }
q_2-q_1\to 0^+.
\]
At the moment of overtaking, however, the presence of the factor
$\wid{\theta}_{\frac12}(q_j-q_i)$ will decrease the speed of the
cliffon \# 1 of a factor $p_2$, while increasing the speed of the
cliffon \# 2 of a factor $p_1$, so that the two cliffons merge
into a single one, traveling with the sum of their "initial"
speeds. This behavior is depicted in Figure 2, as seen in the
reference frame of the rightmost cliffon.
\subsection*{Acknowledgments} 
The author wishes to thank B.
Dubrovin, T. Grava, P. Lorenzoni, and Y. Zhang for useful
discussions. He is also grateful to the organizers of SPT04 (Cala
Gonone --IT, May 2004), of the Workshop  {\em
Analytic and Geometric theory of the Camassa-Holm equation and Integrable
Systems}, (Bologna--IT, September 2004),
and of the GNFM annual Meeting (Montecatini
-- IT, October 2004) for the possibility of presenting a few of
the results herewith collected in those Conferences. This work has
been partially supported by INdAM-GNFM under the research project
{\em Onde nonlineari, struttura $\tau$ e geometria delle variet\`a
invarianti: il caso della gerarchia di Camassa-Holm}, by the ESF
project {\em MISGAM}, by the by the Italian M.I.U.R. under the
research project {\em Geometric methods in the theory of nonlinear
waves and their applications}, and by the European Community
through the FP6 Marie Curie RTN {\em ENIGMA} (Contract number
MRTN-CT-2004-5652).

\end{document}